\documentclass[a4paper]{jpconf}
\usepackage{graphicx}
\usepackage{float}
\begin{document}
\title{The New Compact 341 Model: Higgs Decay Modes}
\author{N.Mebarki, M.Djouala, J.Mimouni and H.Aissaoui}

\address{Laboratoire de Physique Mathématique et Subatomique\\
	Mentouri University, Constantine1, Algeria}

\ead{nnmebarki@yahoo.fr}

\begin{abstract}
New developments in the anomaly free compact 341 model are discussed and the
higgs bosons decay modes are studied taking into account the contributions of new fermions,
gauge bosons and scalar bosons predicted by the model. It is shown from signal strengths and
the branching ratios of the various decay modes analysis and the LHC constraints that there is a
room for this extended BSM model and it is viable. 
\end{abstract}
\section{Introduction}
\label{intro}
Despite all successes of the standard model, many questions remained unsolved and not well
understood like dark matter, neutrinos oscillation, matter anti-matter asymmetry etc... Trying to find a solution to those problems, one needs to extend the standard model and go beyond (BSM). The most proposed model on the literature are the ones with two-Higgs doublets (THDM)\cite{1}, supersymmetry \cite{2}, 331, extra dimensions \cite{3} and 341 gauge models \cite{4,11}.\\
\indent Among those extensions, we focus on a model which is based on the
$SU(3)_C \otimes SU(4)_L \otimes U(1)_N$ gauge symmetry (denoted by 341 model for a short hand). This model has new particles like exotic
quarks, new gauge bosons $K_{0}$, $K_{0}^{\prime}$, $K_{1}^{\mp}$, $X^{\mp}$, $V^{\mp\mp}$, $Y^{\mp}$, $Z_{0}^{\prime}$ and $Z^{\prime\prime}$. Moreover, the 341 model has a very
specific arrangement of the fermions into generations; for leptons, one has both right and left handed helecities arranged in the same multiplet. In order to make the model anomaly free, the second and
(4) third quarks families has to belong to the conjugate $4^{*}$ fundamental representation of the gauge group, while the first family transforms as a quadruplet in the fundamental representation. In
this compact 341 model, we have a minimum of three scalars quartets\cite{8} and after SSB which is achieved via three steps, one ends up with three CP even neutral higgses $h_{1}$ , $h_{2}$ and $h_{3}$ and eight CP
odd massive higgses $h_{1}^{\mp}$, $h_{2}^{\mp}$ and $h^{\mp\mp}$.\\
\indent In this paper, we focus on the analysis of the neutral Higgs decays modes and discuss the signal strengths and the branching ratios of the various decay modes as well as the LHC constraints and show that there is a room for this extended BSM model and it is viable. In section \ref{model}, we present a brief review of the theoretical model. In section \ref{Higgs}, we give the various analytical expressions of the partial decays width which we have derived using the new Feynman rules of the model. Finally, in section \ref{Numerical results and conclusions}, we give our numerical results concerning the signal strength of the various higgses branching ratios, after imposing the self consistency and compatibility constraints on the scalar potential of the model
like triviality, unitarity, vacuum stability and non-ghost conditions, make comparison with the signal strengths of the recent experimental data reported by ATLAS, CMS and combined ATLAS+CMS and
draw our conclusions.
\section{The theoretical model}
\label{model}
The gauge group structure of the model is $SU(3)_C \otimes SU(4)_L \otimes U(1)_N$ and the electric charge operator $\widehat{Q}$ is defined as \cite{9}:
\begin{equation}\label{eq:So}
\widehat{Q}/e=\frac{1}{2}\bigg(T_{3}-\frac{1}{\sqrt{3}} T_{8}-\frac{4}{\sqrt{6}}\gamma \bigg)+N
\end{equation}
where:
\begin{equation}
\lambda_{3} =diag(1,-1,0,0)\qquad \qquad \lambda_{8} =
\frac{1}{\sqrt{3}}diag(1,1,-2,0) \qquad
\lambda_{15} = \frac{1}{\sqrt{6}}diag(1,1,1,-3).
\end{equation}
The fermions content of this model is as follows \cite{4}: for the leptons (resp. quarks) denoted by $L_{a L}$ and $Q_{1L}$, $Q_{iL}$ respectively one has,
$$
f_{aL}= \left(
\begin{array}{ccc}
\nu_{a} \\
l_{a} \\
\nu_{a}^{c}\\
l_{a}^{c} \\
\end{array}
\right)\sim (1,4,0),~~~
Q_{1L}= \left(
\begin{array}{ccc}
u_{1} \\
d_{1} \\
U_{1}\\
J_{1} \\
\end{array}
\right)\sim (3,4,\frac{2}{3}),~~~Q_{iL}= \left(
\begin{array}{ccc}
d_{i} \\
u_{i} \\
D_{i}\\
J_{i} \\
\end{array}
\right)\sim (3,4^{*},\frac{-1}{3})
$$
Where a=1,2,3 and i=2,3. Here 
$U_{1}$, $J_{1}$, $D_{i}$ and
$J_{i}$ are exotic quarks with electric charges
$\frac{2}{3}$, $\frac{5}{3}$, $\frac{-1}{3}$
and $\frac{-4}{3}$ respectively. Right-handed quarks transform as
$u_{1R}$$(3,1,\frac{2}{3})$, $d_{1R}$
$(3,1,\frac{-1}{3})$, $U_{1R}$
$(3,1,\frac{2}{3})$, $J_{1R}$$(3,1,\frac{5}{3})$,
$u_{iR}$ $(3,1,\frac{2}{3})$, $d_{iR}$
$(3,1,\frac{-1}{3})$, $D_{iR}$
$(3,1,\frac{-1}{3})$, $J_{iR}$$ (3,1,\frac{-4}{3})$. The most
general scalar potential with a $Z_{3}$
discrete symmetry in the compact 341 model is given by \cite{4}:
\begin{eqnarray}
V(\eta,\rho,\chi)&=&\mu_{\eta}^{2}\eta^{\dag}\eta+\mu_{\rho}^{2}\rho^{\dag}\rho+\mu_{\chi}^{2}\chi^{\dag}\chi+\lambda_{1}(\eta^{\dag}\eta)^{2}+\lambda_{2}(\rho^{\dag}\rho)^{2}+\lambda_{3}(\chi^{\dag}\chi)^{2}\nonumber \\
&+&\lambda_{4}(\eta^{\dag}\eta)(\rho^{\dag}\rho)+\lambda_{5}(\eta^{\dag}\eta)(\chi^{\dag}\chi)+\lambda_{6}(\rho^{\dag}\rho)(\chi^{\dag}\chi)+\lambda_{7}(\rho^{\dag}\eta)(\eta^{\dag}\rho)
\nonumber
\\&+&\lambda_{8}(\chi^{\dag}\eta)(\eta^{\dag}\chi)+\lambda_{9}(\rho^{\dag}\chi)(\chi^{\dag}\rho),
\end{eqnarray}
Where $\mu^{2}_{\mu\rho\chi}$ are the mass dimension parameters and
$\lambda_{S}$ S=$\overline{1,9}$ are dimensionless coupling constants.
The scalars quadruplets $\eta$, $\rho$ and $\chi$ (which are necessary to generate masses) are given by the following
quartets:
$$
\eta= \left(
\begin{array}{ccc}
\eta_{1}^{0}\\
\eta_{1}^{-} \\
\eta_{2}^{0} \\
\eta_{+}^{2}
\end{array},
\right)=\left(
\begin{array}{ccc}
\frac{1}{\sqrt{2}}(R_{\eta_{1}}+iI_{\eta_{1}})\\
\eta_{1}^{-} \\
 \frac{1}{\sqrt{2}}(v_{\eta_{2}}+R_{\eta_{2}}+iI_{\eta_{2}})\\
\eta_{2}^{+}
\end{array},
\right)\sim (1,4,0),
$$
$$ \rho=\left(
\begin{array}{ccc}
\rho_{1}^{+}\\
\rho^{0} \\
\rho_{2}^{+} \\
\rho^{++}
\end{array},
\right)=\left(
\begin{array}{ccc}
\rho_{1}^{+}\\
 \frac{1}{\sqrt{2}}(v_{\rho}+R_{\rho}+iI_{\rho}) \\
\rho_{2}^{+} \\
\rho^{++}
\end{array},
\right)\sim(1,4,1),
$$
$$
\chi=
\left(
\begin{array}{ccc}
\chi_{1}^{-}\\
\chi^{--} \\
\chi_{2}^{-} \\
\chi^{0}
\end{array}
\right)\left(
\begin{array}{ccc}
\chi_{1}^{-}\\
\chi^{--} \\
\chi_{2}^{-} \\
\frac{1}{\sqrt{2}}(v_{\chi}+R_{\chi}+iI_{\chi})
\end{array},
\right)\sim (1,4,-1).
$$
The reason to choose the $\eta$ quadruplet developing VeV only in the
only in the $3^{rd}$ component is to avoid mixings
between ordinary quarks and exotic ones. Imposing the tadpole conditions:
\begin{eqnarray}
	\mu^{2}_{1}+\lambda_{1}v_{\eta}^{2}+\frac{1}{2}\lambda_{4}v_{\rho}^{2}+\frac{1}{2}\lambda_{5}v_{\chi}^{2}=0,\qquad
		\mu^{2}_{2}+\lambda_{3}v_{\rho}^{2}+\frac{1}{2}\lambda_{4}v_{\eta}^{2}+\frac{1}{2}\lambda_{6}v_{\chi}^{2}=0\nonumber
		\end{eqnarray}
	\begin{equation}
		\mu^{2}_{3}+\lambda_{3}v_{\chi}^{2}+\frac{1}{2}\lambda_{5}v_{\eta}^{2}+\frac{1}{2}\lambda_{6}v_{\rho}^{2}=0\\\nonumber	
\end{equation}
helps to find the CP-even neutral scalars mass matrix in the basis ($R_{\rho},R_{\chi},R_{\eta}$ , whose eigenvalues are \cite{7}:
\begin{eqnarray}
M^{2}_{h_{1}}=
\lambda_{2}\upsilon_{\rho}^{2}+\frac{\lambda_{3}\lambda^{2}_{4}+\lambda_{6}(\lambda_{1}\lambda_{6}-\lambda_{4}\lambda_{5})}{\lambda^{2}_{5}-4\lambda_{1}\lambda_{3}}\upsilon_{\rho}^{2},~~
M_{h_{2}}^{2}=
c_{1}\upsilon_{\chi}^{2}+c_{2}\upsilon_{\rho}^{2}\approx c_{1}\upsilon_{\chi}^{2},~~
M_{h_{3}}^{2}= c_{3}\upsilon_{\chi}^{2}+c_{4}\upsilon_{\rho}^{2}\approx c_{3}\upsilon_{\chi}^{2}.
\end{eqnarray}
representing the masses of the physical scalars $h_{1}$, $h_{2}$ and $h_{3}$
respectively (the lightest neutral scalar
$h_{1}$ is identified as SM like Higgs boson) and eigenstates:
\begin{equation}
h_{1}=R_{\rho},~~h_{2}=aR_{\chi}+bR_{\eta_{2}},~~h_{3}=cR_{\chi}+dR_{\eta_{2}}
\end{equation}
where 
\begin{eqnarray}
a=\frac{\lambda_{1}-\lambda_{3}-\sqrt{(\lambda_{1}-\lambda_{3})^{2}+\lambda_{5}^{2}}}{\lambda_{5}^{2}+(\lambda_{1}-\lambda_{3}-\sqrt{(\lambda_{1}-\lambda_{3})^{2}+\lambda_{5}^{2}})},~~
b=\frac{\lambda_{5}}{\lambda_{5}^{2}+(\lambda_{1}-\lambda_{3}-\sqrt{(\lambda_{1}-\lambda_{3})^{2}+\lambda_{5}^{2}})}\\
c=\frac{\lambda_{1}-\lambda_{3}+\sqrt{(\lambda_{1}-\lambda_{3})^{2}+\lambda_{5}^{2}}}{{\lambda_{5}^{2}+(\lambda_{1}-\lambda_{3}+\sqrt{(\lambda_{1}-\lambda_{3})^{2}+\lambda_{5}^{2}})}},~~
d=\frac{\lambda_{5}}{{\lambda_{5}^{2}+(\lambda_{1}-\lambda_{3}+\sqrt{(\lambda_{1}-\lambda_{3})^{2}+\lambda_{5}^{2}})}}
\end{eqnarray}
The SSB steps are:
\begin{eqnarray}
SU(4)_L\otimes U(1)_N \longrightarrow^{v_{\chi}}SU(3)_L\otimes U(1)_X, SU(3)_L\otimes U(1)_X \longrightarrow^{v_{\eta}}SU(2)_L\otimes U(1)_Y\nonumber,
\end{eqnarray}
\begin{eqnarray}
 SU(2)_L\otimes U(1)_Y \longrightarrow^{v_{\rho}} U(1)_{QED}.\nonumber
\end{eqnarray}
Here stands for the weak isospin quantum number and the VeV’s are such that $\sim$ 246 GeV, $v_{\eta}\sim \mathcal{O}$(TeV) and $v_{\chi}\sim \mathcal{O}$(TeV) ($v_{\chi}\sim v_{\eta}$) The masses of the charged gauge bosons in this model
are:
\begin{eqnarray}
M_{W{\mp}}^{2}&=&\frac{g^{2}_{L}}{4}\upsilon^{2}_{\rho},
M_{K^{0},K^{0'}}^{2}=\frac{g^{2}_{L}}{4}\upsilon^{2}_{\eta},
M_{K_{1}^{\mp}}^{2}=\frac{g^{2}_{L}}{4}(\upsilon^{2}_{\eta}+\upsilon^{2}_{\rho}),
M_{X^{\mp}}^{2}=\frac{g^{2}_{L}}{4}\upsilon^{2}_{\chi},
 M_{V^{\mp\mp}}^{2}=\frac{g^{2}_{L}}{4}(\upsilon^{2}_{\rho}+\upsilon^{2}_{\chi}),\nonumber
\end{eqnarray}
\begin{equation}
	M_{Y^{\mp}}^{2}=\frac{g^{2}_{L}}{4}(\upsilon^{2}_{\eta}+\upsilon^{2}_{\chi}).
\end{equation}
and of the neutral ones:
\begin{eqnarray}
M_{\gamma}=0,
M_{Z}^{2}=\frac{g^{2}\upsilon_{\rho}^{2}}{4c_{W}^{2}},~~
M_{Z'}^{2}=\frac{g^{2}c_{W}^{2}\upsilon_{\eta}^{2}}{h_{W}},~~ M_{Z''}^{2}=\frac{g^{2}\upsilon_{\eta}^{2}\bigg((1-4s_{W}^{2})^{2}+h_{W}^{2}\bigg)}{8h_{W}(1-4s_{W}^{2})}.	
\end{eqnarray}
where
\begin{eqnarray}
W^{\mp}&=&\frac{(W_{\mu}^{1}\mp
	iW_{\mu}^{2})}{\sqrt{2}},
K^{0},K^{'0}=\frac{(W_{\mu}^{4}\mp iW_{\mu}^{5})}{\sqrt{2}},
K^{\mp}_{1}=\frac{(W_{\mu}^{6}\mp iW_{\mu}^{7})}{\sqrt{2}}
,
\end{eqnarray}
and
\begin{eqnarray}
X^{\mp}=\frac{(W_{\mu}^{9}\mp iW_{\mu}^{10})}{\sqrt{2}},
V^{\mp\mp}&=&\frac{(W_{\mu}^{11}\mp iW_{\mu}^{12})}{\sqrt{2}},
Y^{\mp}=\frac{(W_{\mu}^{13}-iW_{\mu}^{14})}{\sqrt{2}}.
\end{eqnarray}
Here, $\cos_{\theta}=c_{w}$, $\sin_{\theta}=s_{w}$ and $h_W=3-4s_{w}^{2}$.
It is worth to mention that, the most attractive phenomenological features of the model is that in
addition to the reproduction of all phenomenological success of SM, it has only 03 families of quarks
and leptons and computation of the of U(1) gauge group running coupling shows the presence of a
Landau pole at a scale around 5 TeV. This implies the existence of a natural cut off for the model
around the TeV scale and therefore solving hierarchy problem. Moreover, this cutoff can be used to implement fermions masses that are not generated by Yukawa couplings including neutrinos masses
and consequently one has a natural Dark matter candidate.
\section{Higgs decays modes in the compact 341 model}
\label{Higgs}
\begin{figure}[H]%
	\begin{center}
	\includegraphics[width=.4\textwidth]{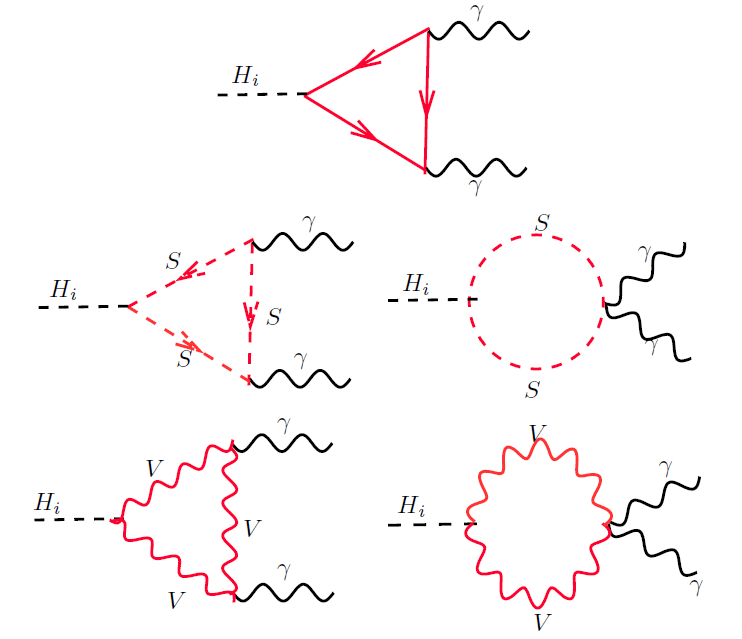}
	\caption{\label{fig:1i} the one-Loop diagrams contributing to $h\longrightarrow \gamma\gamma$
		decay modes.}
	\end{center}
\end{figure}

To determine the different SM Higgs-like branching ratios, we have derived all the Feynman rules of
the various vertices within the compact 341 model \cite{7,8} and get explicit analytical expressions of the
various $h_{1}$ partial decay widths channels. The main Feynman diagrams contributing to the neutral
Higgs (denoted by in fig.\ref{fig:1i}) double photon production ($h_{1}\longrightarrow \gamma\gamma$)are displayed in Fig.\ref{fig:1i}.
The partial decay width is shown to have the following form:
\begin{eqnarray}
\Gamma(h_{1}\longrightarrow
\gamma\gamma)&=&\frac{\alpha^{2}m_{H}^{3}}{1024\pi^{3}}|\sum_{V}\frac{g_{HVV}}{m^{2}_{V}}Q_{V}^{2}A_{1}(\tau_{V})+\sum_{f}\frac{2g_{Hff}}{m_{f}}Q_{f}^{2}N_{c,f}A_{\frac{1}{2}}(\tau_{f})\nonumber
\\&+&\sum_{f}\frac{g_{HSS}}{m_{S}^{2}}Q_{S}^{2}N_{c,S}A_{0}(\tau_{S})|^{2}
\end{eqnarray}
Straightforward but lengthy calculations using the new derived Feynman vertices leads also to:
\begin{eqnarray}
\Gamma_{341}(h_{1}\longrightarrow
\overline{l}l)&=&\frac{g^{2}}{32\pi}\frac{m_{l}^{2}}{m_{W}^{2}}m_{h_{1}}\bigg(1-\frac{4m^{2}_{l}}{m^{2}_{h_{1}}}\bigg)^{\frac{3}{2}},\nonumber\\
\Gamma_{341}(h_{1}\longrightarrow
\overline{b}b)&=&\frac{3g^{2}}{32\pi}\frac{m_{b}^{2}}{m_{W}^{2}}m_{h_{1}}\bigg(1-\frac{4m^{2}_{b}}{m^{2}_{h_{1}}}\bigg)^{\frac{3}{2}},\nonumber\\
\Gamma(h_{1}\longrightarrow \gamma
Z)&=&\frac{\alpha^{2}m_{h_{1}}^{3}}{512\pi^{3}}\bigg(1-\frac{M^{2}_{Z}}{M^{2}_{h_{1}}}\bigg)^{3}\bigg|\frac{2}{\upsilon}\frac{\mathcal{A}_{SM}}{\sin\theta_{W}}+\mathcal{A}\bigg|^{2},\nonumber\\
\Gamma_{341}(h_{1}\longrightarrow
W^{*}W)&=&\frac{3g^{4}m_{h_{1}}}{512\pi^{3}}F\bigg(\frac{m_{W}}{m_{h_{1}}}\bigg)\nonumber,\\
\Gamma_{341}(h_{1}\longrightarrow Z^{*}Z)&=&
\frac{g^{4}g^{2}_{h_{1}ZZ}}{2048\pi^{3}C_{W}^{2}}m_{h_{1}}F\bigg(\frac{m_{Z}}{m_{h_{1}}}\bigg)\bigg(\sum_{j=quarks}(g_{jV}^{2}+g_{jA}^{2})
+\sum_{l=leptons}(g_{lV}^{2}+g_{lA}^{2})\bigg)\nonumber\\.
\end{eqnarray}
where  $\tau_i=4m^{2}_{i}/m^{2}_{h_{1}}$, V,f and S refer to Spin1, Spin$\frac{1}{2}$ and Spin0 particles respectively. The loop
functions are given by :\newpage
\begin{eqnarray}
A_{1}(x)&=&-x^{2}\bigg(2x^{-2}+3x^{-1}+3(2x^{-1}-1)f(x^{-1})\bigg),~~
A_{\frac{1}{2}}(x)=2x^{2}\bigg(x^{-1}+(x^{-1}-1)f(x^{-1})\bigg),\nonumber
\end{eqnarray}
\begin{eqnarray}
A_{0}(x)&=&-x^{2}\bigg(x^{-1}-f(x^{-1})\bigg).
\end{eqnarray}
with 
\begin{equation}
\label{1-1} f(x)=\left\lbrace
\begin{array}{ll}\bigskip
\arcsin^{2}\sqrt{x}~~~~~for~~ x\geq 1,\\
\frac{-1}{4}\bigg(\ln(\frac{1+\sqrt{1-x^{-1}}}{1-\sqrt{1-x^{-1}}
})-\imath\pi\bigg)^{2}~~~~~for ~~ x< 1.
\end{array}
\right.
\end{equation}
Similarly,
\begin{eqnarray}
\mathcal{A}&=&\frac{g_{h_{1}VV}}{m^{2}_{V}}g_{ZVV}\tilde{\mathcal{A}}_{1}(\tau_{V},\lambda_{V})+\widetilde{N}_{c,f}\frac{4N_cQ_{f}}{m^2_{f}}g_{h_{1}ff}(g^{L}_{Zff}
+g^{R}_{Zff})\tilde{\mathcal{A}}_{\frac{1}{2}}(\tau_{f},\lambda_{f})\nonumber\\
&-&\frac{2N_cQ_S}{m^{2}_{S}}g_{h_{1}SS}g_{ZSS}\tilde{\mathcal{A}}_{0}(\tau_{S},\lambda_{S}),
\end{eqnarray}
with 
\begin{eqnarray}
\tilde{A}_{1}(x,y)&=&4(3-\tan^{2}\theta_{W})I_{2}(x,y)+\bigg((1+2x^{-1})\tan^{2}\theta_{W}-(5+2x^{-1})\bigg)I_{1}(x,y),\nonumber\\
\tilde{A}_{\frac{1}{2}}(x,y)&=&I_{1}(x,y)-I_{2}(x,y),~~
\tilde{A}_{0}(x,y)=I_{1}(x,y),
\end{eqnarray}
where
\begin{eqnarray*}
I_{1}(x,y)&=&\frac{xy}{2(x-y)}+\frac{x^{2}y^{2}}{2(x-y)^{2}}\bigg(f(x^{-1})-f(y^{-1})\bigg)+\frac{x^{2}y}{(x-y)^{2}}\bigg(g(x^{-1})-g(y^{-1})\bigg),\\
I_{2}(x,y)&=&\frac{-xy}{2(x-y)}\bigg(f(x^{-1})-f(y^{-1})\bigg).
\end{eqnarray*}
\begin{eqnarray*}
\label{1-1} g(x)=\left\lbrace
\begin{array}{ll}\bigskip
\sqrt{x^{-1}-1}\arcsin\sqrt{x}~~~~for~~~~ x\geq 1,\\
\frac{\sqrt{1-x^{-1}}}{2}\bigg(\ln(\frac{1+\sqrt{1-x^{-1}}}{1-\sqrt{1-x^{-1}}
})-\imath\pi\bigg)~~~~for ~~~~x< 1.
\end{array}
\right.
\end{eqnarray*}
and
\begin{eqnarray}
F(x)&=&-|1-x^{2}|(\frac{47}{2}x^{2}-\frac{13}{2}+\frac{1}{x^{2}})
-\frac{3}{2}(1-6x^{2}+4x^{4})\ln(x)
+\frac{3(1-8x^{2}+20x^{4})}{\sqrt{4x^{2}-1}}\arccos(\frac{3x^{2}-1}{2x^{3}}).\nonumber\\
\end{eqnarray}
Here $\mathcal{A}_{SM}$ represents the SM contribution, $\lambda_{i}=4m^{2}_{i}/m^{2}_{Z}$ and $g_{h_{1}VV}$,$g_{ZVV}$,$g_{h_{1}ff}$,$g^{L}_{Zff}$,$g^{R}_{Zff}$,$g_{h_{1}SS}$, $g_{ZSS}$, $g_{h_{1}VV}$ are couplings constants. Here, $Q_{V}$, $Q_{f}$,$Q_{S}$ are electric charges of the vectors,
fermions and scalars and $N_{c;f}$ ; $N_{c;S}$ are the number of fermion and scalar colors respectively\cite{12}. It is
Here very important to mention that the SM contribution of the diphoton decay channel comes
essentially from the one loop top quark and the gauge bosons W. However, in the 341 Model, beside the W and the top quark, it includes the new heavy gauge bosons $K_{1}^{\mp}$ and $V^{\mp\mp}$, and the
charged higgs bosons $h_{1}^{\mp}$ , $h_{2}^{\mp}$ and $h^{\mp\mp}$ (there is no direct coupling between the exotic quarks
and the Higgs like-boson $h_{1}$). Regarding $h_{2}$ and $h_{3}$ higgs bosons, the expressions of most of
the various decay widths are the same as the ones of the higgs $h_{1}$ except that the couplings are different and replace $m_{h_{1}}$ by $m_{h_{2}}$ or $m_{h_{3}}$. Among the interesting new decay modes, one has $h_{2}\longrightarrow h_{1}h_{1}$, $h_{3}\longrightarrow h_{1}h_{1}$, $h_{3}\longrightarrow h_{2}h_{2}$ and $h_{3}\longrightarrow h_{2}h_{1}$ with the corresponding decay width:
\begin{eqnarray*}
\Gamma_{341}(h_{2}\longrightarrow h_{1}h_{1})&=&\frac{1}{16\pi
	m_{h_{2}}}(g_{h_{2}h_{1}h_{1}})^{2}\bigg(1-\frac{4m_{h_{1}}^{2}}{m^{2}_{h_{2}}}\bigg)^{\frac{1}{2}},\nonumber\\
\Gamma_{341}(h_{3}\longrightarrow h_{1}h_{1})&=&\frac{1}{16\pi
	m_{h_{3}}}(g_{h_{3}h_{1}h_{1}})^{2}\bigg(1-\frac{4m_{h_{1}}^{2}}{m^{2}_{h_{3}}}\bigg)^{\frac{1}{2}},\nonumber\\
\Gamma_{341}(h_{3}\longrightarrow h_{2}h_{2})&=&\frac{1}{16\pi
	m_{h_{3}}}(g_{h_{3}h_{2}h_{2}})^{2}\bigg(1-\frac{4m_{h_{2}}^{2}}{m^{2}_{h_{3}}}\bigg)^{\frac{1}{2}},
\end{eqnarray*}
\begin{equation}
\Gamma_{341}(h_{3}\longrightarrow h_{2}h_{1})=\frac{1}{16\pi
	m_{h_{3}}^{2}}(g_{h_{3}h_{2}h_{1}})^{2}\bigg(\frac{m_{h_{2}}^{4}}{m_{h_{3}}^{2}}-\frac{2m_{h_{1}}^{2}m_{h_{2}}^{2}}{m^{2}_{h_{3}}}-2m_{h_{2}}^{2}+\frac{m_{h_{1}}^{4}}{m_{h_{3}}^{2}}-2m_{h_{1}}^{2}+m_{h_{3}}^{2}\bigg)^{\frac{1}{2}},
\end{equation}
where 

\begin{eqnarray*}
g_{h_{2}h_{1}h_{1}}=v_{\chi}\bigg(\frac{\lambda_{6}}{2}\gamma+\frac{\lambda_{4}}{2}\frac{v_{\eta}}{v_{\chi}}\alpha\bigg),~
g_{h_{3}h_{1}h_{1}}=v_{\chi}\bigg(\frac{\lambda_{6}}{2}\sigma+\frac{\lambda_{4}}{2}\frac{v_{\eta}}{v_{\chi}}\beta\bigg).
\end{eqnarray*}
\begin{equation}
g_{h_{3}h_{2}h_{2}}=\frac{\lambda_{5}}{2}\bigg(\upsilon_{\chi}(\alpha^{2}\sigma+2\alpha\beta\gamma)+\upsilon_{\eta}(\beta\gamma^{2}+2\alpha\gamma\sigma)\bigg),~~
g_{h_{3}h_{2}h_{1}}=\lambda_{4}\upsilon_{\rho}\alpha\beta+\lambda_{6}\upsilon_{\rho}\gamma\sigma.
\end{equation}
The parameters $\alpha$, $\beta$ , $\gamma$ and $\sigma$ are functions of the potential parameters $\lambda$'s (see refs.\cite{7,8}). It is worth
to mention that in the decay modes $h_{2}\longrightarrow \gamma\gamma$
and $h_{2}\longrightarrow Z\gamma$
, one has additional contributions of
exotic fermions , charged scalars and the new gauge bosons (for more details see refs.\cite{7,8}).
\section{Numerical results and conclusions}
\label{Numerical results and conclusions}
\begin{figure}[H]%
	\begin{center}
	\includegraphics[width=.26\textwidth]{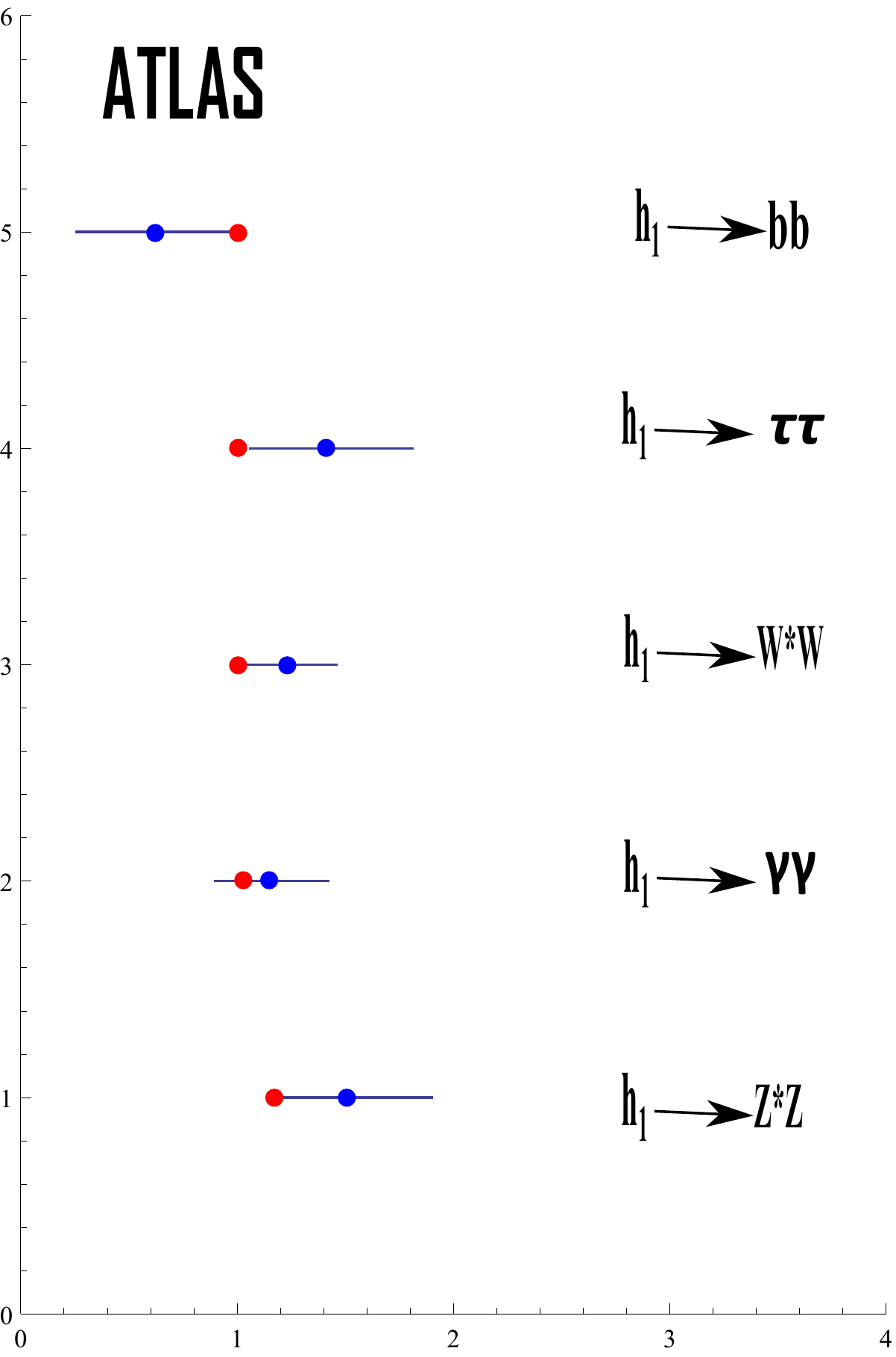}
	~~~~~~~~~~~~~~~~~~~~~~
	\includegraphics[width=.26\textwidth]{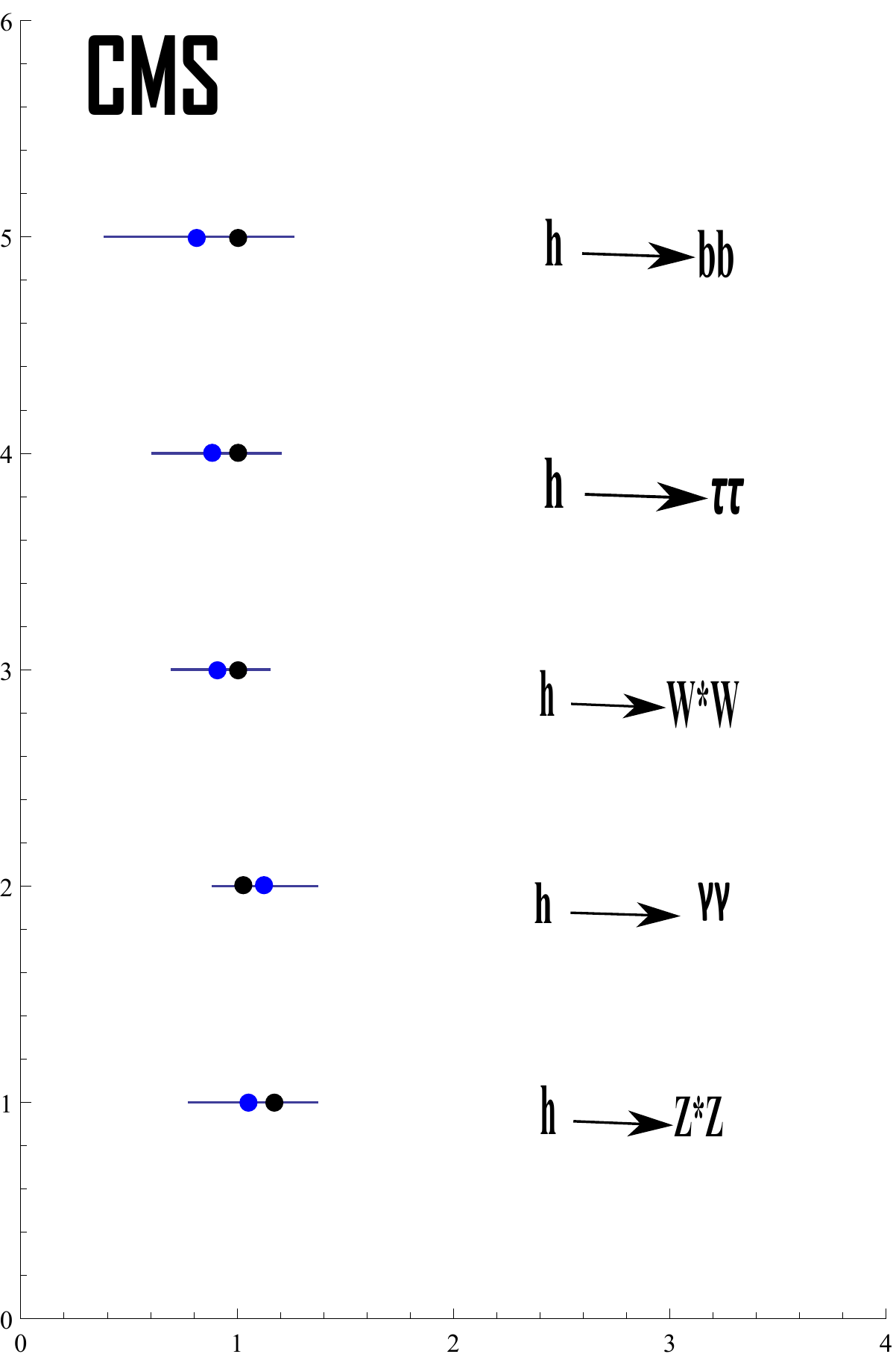}
		\caption{\label{fig:i22} Signal Strengths for various decay modes compared to the ATLAS and CMS run2 data.}
	\end{center}
\end{figure}

\begin{figure}[H]%
	\begin{center}
		\includegraphics[width=.26\textwidth]{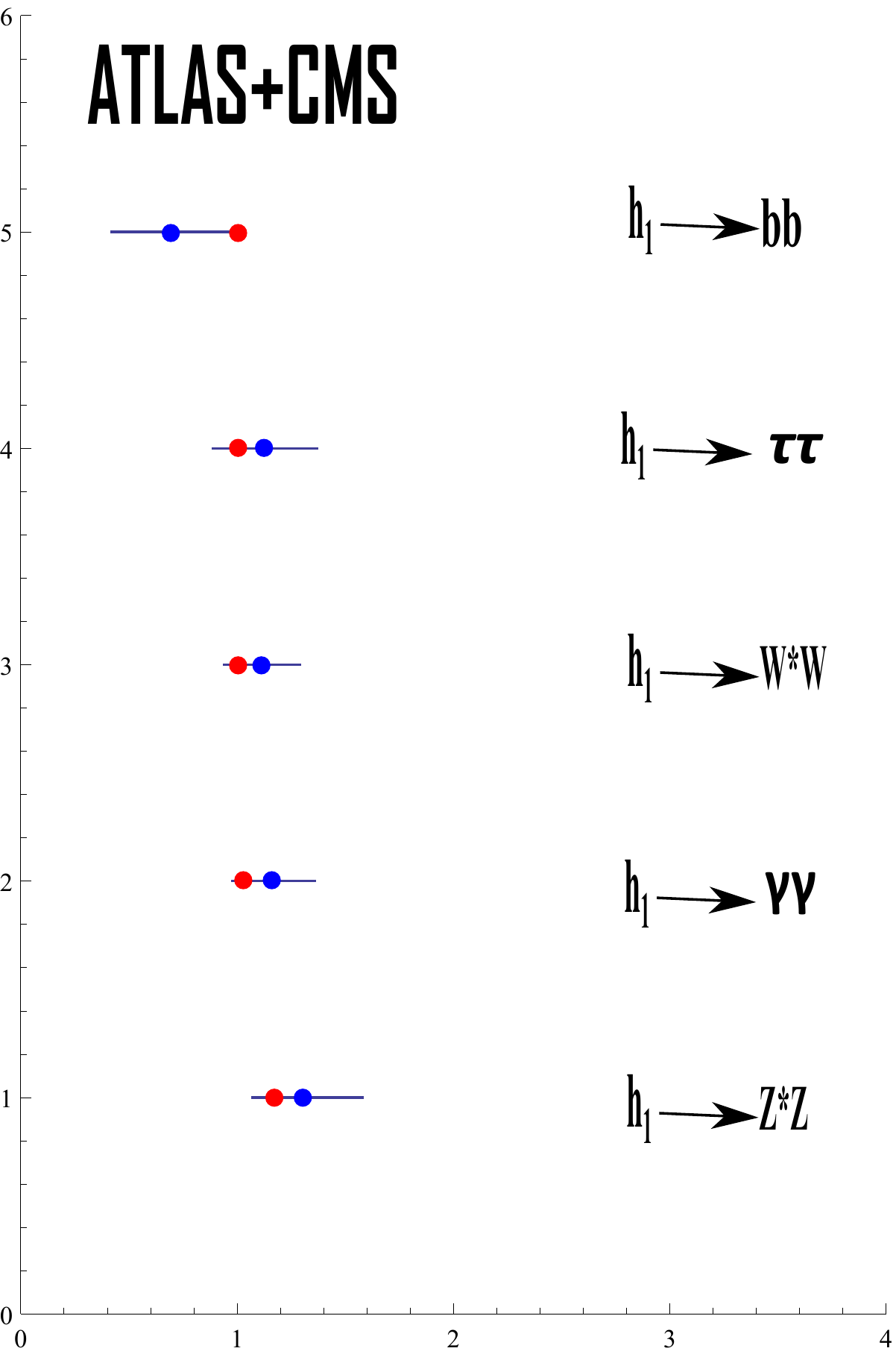}
		\caption{\label{fig:i33} Signal Strengths for various SM like higgs decay modes compared to the combined
			ATLAS-CMS run2 data}
	\end{center}
\end{figure}

We have calculated the signal strength for each individual decay channel in the context of the compact
341 model, an in order to reproduce the experimental results (e.g.
$m_{h_{1}}\sim$ 126 GeV etc...), one
has to take as inputs $v_{\chi}\sim v_{\eta}\sim 2$ TeV
( because of the fact that 341 model has a Landau pole at
a round the TeV scale),
$m_{exotic~quarks}\sim$ 750 GeV
(from the LHC experimental data concerning the
lower bounds on exotic quarks),
$m_{h_{2}}\sim$ 700 GeV
and for gauge bosons see table \ref{tab:aA}. Moreover, other inputs are the scalar potential couplings $\lambda_{i}$'s selected from a random number generator and a Monte Carlo simulation after putting the self consistency constraints. In fact, we have obtained a confidence band due to the variations of the couplings's within the allowed parameter space region after imposing the noghost, perturbative unitarity, triviality and stability conditions. Table \ref{tab:AAAAAA}, shows the results of the signal strength experimental data of ATLAS, CMS, combined ATLAS+CMS and the predictions of the compact 341 model. Figs.\ref{fig:i22} and \ref{fig:i33}, display
the signal strengths for various decay modes compared to the ATLAS, CMS and the combined ATLAS+CMS run2 data \cite{13,14}. Notice that the predictions of the 341 model are fairly good and compatible with the run2 experimental data. This is a confirmation and a proof of the
viability of the 341 BSM model.
\begin{table}[h]
\caption{\label{tab:aA} Masses of gauge bosons in compact 341 model.}
\begin{center}
\begin{tabular}{llllllll}
\br
Gauge boson&Mass TeV\\
\mr
Z&0.091\\
$Z^{\prime}$&0.79\\
$Z^{\prime\prime}$&2.2\\
$W^{\mp}$&0.08\\
$K^{0},K^{\prime0}$&0.65\\
$K_{1}^{\mp}$&0.655\\
$X^{\mp}$&0.655\\
$V^{\mp\mp}$&0.655\\
$Y^{\mp}$&0.92\\
 \br
\end{tabular}
\end{center}
\end{table}

\begin{table}[H]
	\caption{\label{tab:AAAAAA} Signal strength data of ATLAS,CMS, ATLAS+ CMS and compact 341 model.}
	\begin{center}
		\begin{tabular}{llllllll}
			\br
			Decay channel & ATLAS & CMS & ATLAS+CMS &  The compact 341 model\\
			\mr
		$\mu^{\gamma\gamma}$&$1.15^{+0.27}_{-0.25}$&$1.12^{+0.25}_{-0.23}$&$1.16^{+0.20}_{-0.18}$&1.03\\
		$\mu^{ZZ}$&$1.51^{+0.39}_{-0.34}$&$1.05^{+0.32}_{-0.27}$&$1.31^{+0.27}_{-0.24}$&1.17\\
		$\mu^{WW}$&$1.23^{+0.23}_{-0.21}$&$0.91^{+0.24}_{-0.21}$&$1.11^{+0.18}_{-0.17}$&0.99\\
		$\mu^{\tau\tau}$&$1.41^{+0.40}_{-0.35}$&$0.89^{+0.31}_{-0.28}$&$1.12^{+0.25}_{-0.23}$&0.99\\
		$\mu^{bb}$&$0.62^{+0.37}_{-0.36}$&$0.81^{+0.45}_{-0.42}$&$0.69^{+0.29}_{-0.27}$&0.99\\
		 \br
	\end{tabular}
\end{center}
\end{table}

\begin{figure}[H]%
	\begin{center}
		\includegraphics[width=.7\textwidth]{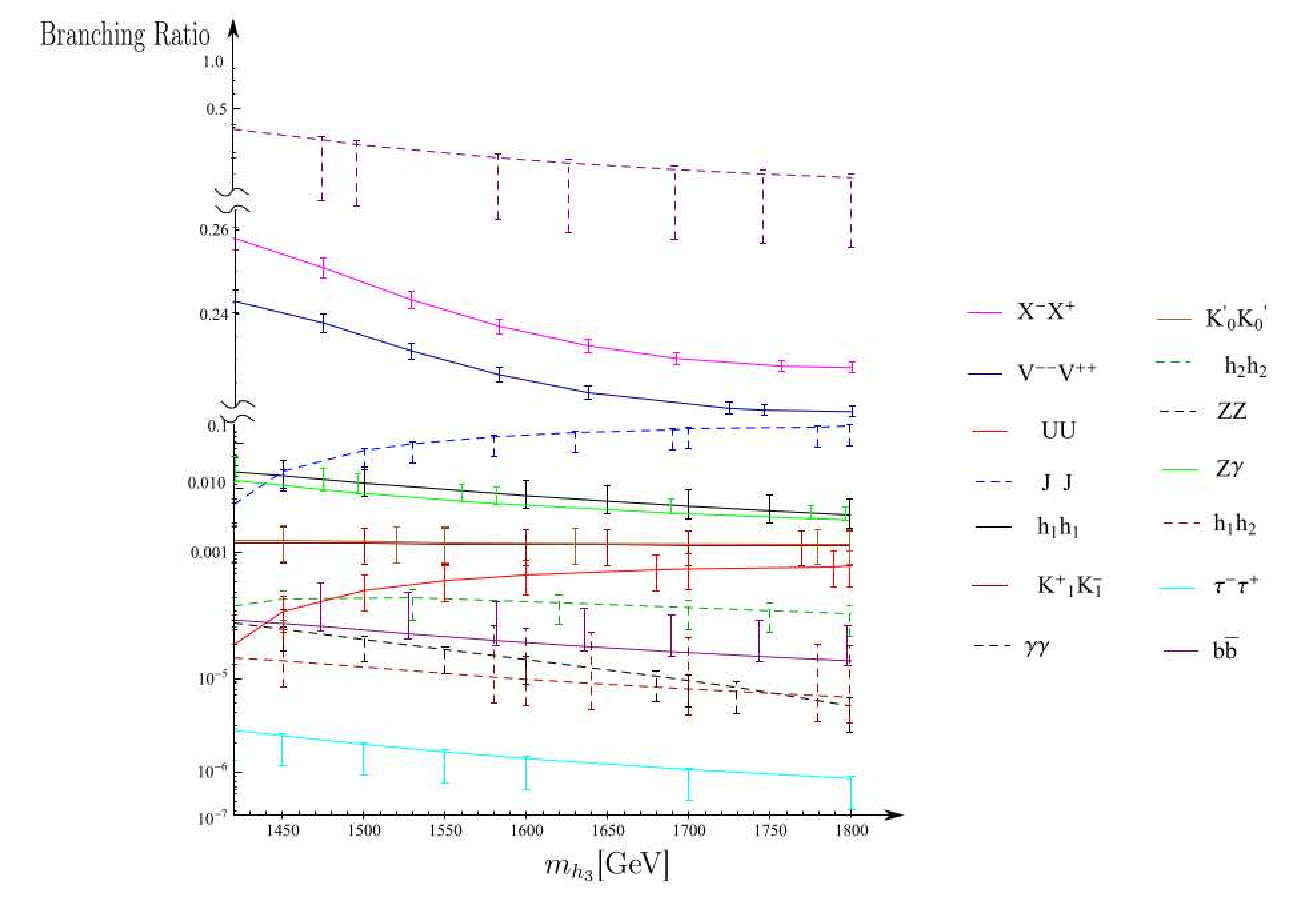}
		\caption{\label{fig:i333} The various $h_{3}$-decay channels in the 341 model}
	\end{center}
\end{figure}

\begin{figure}[H]%
	\begin{center}
		\includegraphics[width=.7\textwidth]{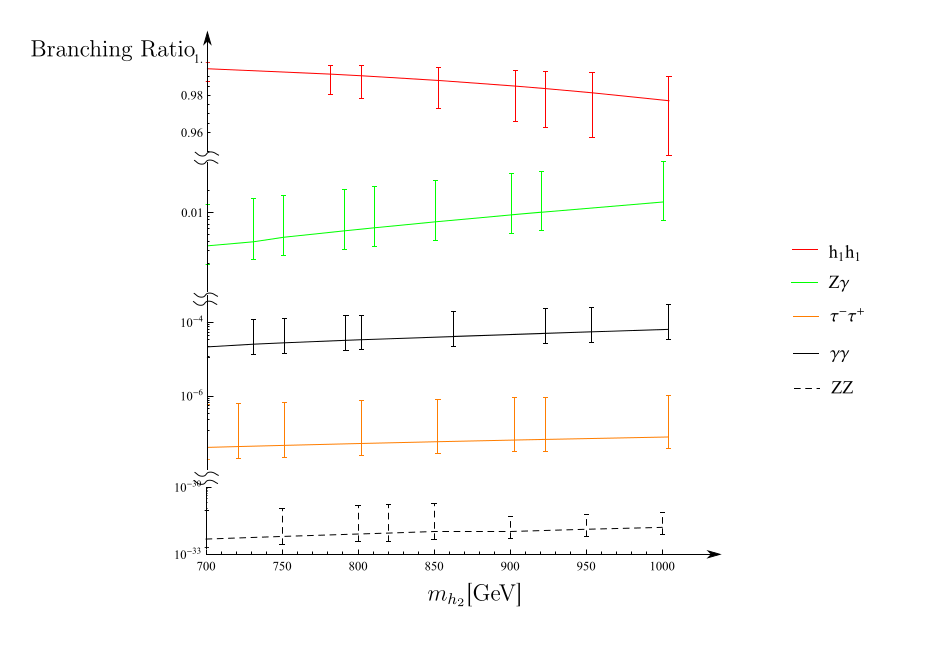}
		\caption{\label{fig:i336}The various $h _{2}$-decay channels in the 341 model}
	\end{center}
\end{figure}

Regarding the heavy higgses $h_2$ and $h_3$ , the branching ratios (BR) for the various mode channels
are shown in figs.\ref{fig:i333} and \ref{fig:i336}. We have used a Monte Carlo simulation taking into account the
theoretical constraints mentioned before. We have checked that there is no big effect regarding
the ambiguity in the choice of the renormalization parameter. For the higgs $h_{2}$ , the dominant decay
mode is $h_{2}\longrightarrow h_{1}h_{1}$ where the branching ratio BR($h_{1}h_{1}$ ) is $\sim$0.98 and it is a decreasing function of
. This could be a good signal for the 341 model regarding the double higgs production
process at the LHC (more study is under investigation) aiming to measure the Higgs self-
coupling and learn about new physics. It is important to mention that measuring the Higgs self
couplings directly probes the structure of the Higgs potential and any deviation of the coupling value
implies BSM physics, also it is important for the vacuum meta stability. Since the Higgs is unstable,
we need to hunt for its decay products like $\bar{b}\tau^{+}\tau^{-}$, $b\bar{b}W^{+}W^{-}$, $\bar{b}b\gamma\gamma$, $\bar{b}b\bar{b}b$
etc..and reconstruct it
from them although the background is very important. To do so and minimize the background, we use
some searches strategies like jet substructure techniques , unboosted and Boosted searches like
exploiting the event kinematic differences between signal and background ,generalize transverse mass cuts to pair production andincrease luminosity etc..in order to gain sensitivity in the main higgs decay
channels then, reconstruct the semi-invisible particle decays and so on. What are the implications of
the di-higgs beyond the standard model physics (BSM) and its relevance to it? how can BSM physics
alter SM di-higgs phenomenology?. It is worth to mention that important resonant and non resonant
enhancements are possible in a large varieties of BSM models. In the compact 341 model, one can
have non resonant enhancement at large transverse momentum due to new loop contributions of
exotic quarks and extra heavy gauge bosons or scalars and/or new (on-shell) resonances like the CP
even higgs $h_2$ where its decay to $h_1 h_1$ is the dominant channel (new states induce large deviations in

inclusive cross section and differential distributions). In this case, one can separate the SM and BSM
contributions using cuts on the invariant mass of the $h_1 h_1$ besides allowing to bound and reconstruct
$\tan\beta=v_{\eta}/v_{\chi}$. For the decay mode $h_{2}\longrightarrow Z\gamma$, one gets a BR(Z$\gamma$)$\sim$0.1, however for $h_{2}\longrightarrow ZZ$
, the branching ratio
BR(ZZ) is very small $\sim 10^{-32}$
and has a big slope as a function of $m_{h_{2}\in}$[0.7,1.0] TeV (see fig.\ref{fig:i333}). For the Higgs h and contrary to h the dominant decay modes are $h_{3}\longrightarrow ZZ$,
$h_{3}\longrightarrow X^{+}X^{-}$ and $h_{3}\longrightarrow V^{++}V^{--}$ where
BR(ZZ) $\in$ [0.12,0.45],
BR$(X^{-}X^{+})\in$ [0.257,0.261] and
BR($V^{++}V^{--})\in$[0.24,0.245] respectively when
$m_{h_{3}}\in$ [0.45,1.8] TeV . However, for the decay
modes $h_{3}\longrightarrow h_{1}h_{1}$ , $h_{3}\longrightarrow$ $J\bar{J}$the branching ratios are $\sim\mathcal{O}(10^{-2})$. Finally, for the $h_{3}\longrightarrow \gamma\gamma$ and ${3}\longrightarrow b\bar{b}$, the branching ratios are very small $\sim\mathcal{O}(10^{-5})$ (for more details see refs.\cite{7}-\cite{8})
\section{Acknowlegments}
We are very grateful to the Algerian Ministry of education and research and DGRSDT for the
financial support.


\section*{References}
\begin{thebibliography}{9}
\bibitem{1} Arhrib A, Benbrik R., Chabab M, Moultaka G and Rahilib L 2012, Higgs boson decay into 2
photons in the type II Seesaw model J. High Energy Phys. JHEP04(2012)136.
\bibitem{2}Aulakh C. S, Melfo A and G. Senjanovic Minimal supersymmetric left-right model Phys.Rev.
D57(1998)4174.
\bibitem {3}Pisano F and Pleitez V 1992,
$SU(4)_L\otimes U(1)_N$ Model for Electroweak Interactions Phys.Rev.D46
(1992)410 arXiv:9206242 [hep-ph].
\bibitem{4} Randall L and Sundrum R 1999, A Large Mass Hierarchy from a Small Extra Dimension Phys.Rev.Lett.83(1999)3370 arXiv:9905221 [hep-ph].
\bibitem{5}Dias A.G., Pinheirob P,R.D., Pires C. A. de S., Rodrigues da Silva P. S. 2014, A compact 341
model at TeV scale Annals of Physics349(2014)232 arXiv:1309.6644v2[hep-ph].
\bibitem{6} Mebarki N 2017 Some Aspects of Higgs Particles in the Compact 341 model Conference,
CoDyCE – LIO International Workshop on Fundamental Theories beyond the Standard
Model 2-4 Oct. 2017.Institut de Physique Nucléaire de Lyon (IPNL) France.
\bibitem{7} Djouala M and Mebarki N 2016 Diphoton Higgs Decay in compact 341 Model, 2èmes Journées
Internationales de Physique Université des Frères Mentouri Constantine 1, Algeria 14 -15
Dec. 2016.
\bibitem{8} Pisano F and Pleitez V 1995
$SU(4)_L \otimes U(1)_N$ Model for the electroweak interactions Phys.
Rev. D51(1995)3865 arXiv:9401272[hep-ph].
\bibitem{9} Cogollo D 2015 Muon Anomalous Magnetic Moment in a
$SU(4)_L \otimes U(1)_N$  Model, Int, Journ.
of Modern Phys. IJMPAV30(2015)1550038 arXiv:1409.8115v3.
\bibitem{10}Carena M, Low I, and Wagner C.E.M 2012 Implications of a Modified Higgs to
Diphoton DecayWidth J. High Energy Phys. JHEP08(2012)060 [arXiv:1206.1082v3].
\bibitem{11} Caetano W, Pires C. A. de S and Rodrigues da Silva P.S 2013 Explaining ATLAS and CMS
Results Within the Reduced Minimal 3-3-1 model, Eur. Phys. J. EPJC73(2013)2607
arXiv:1305.7246v2[hep-ph].
\bibitem{12} ATLAS and CMS Collaborations 2016 Measurements of the Higgs boson production and decay
rates and constraints on its couplings from a combined ATLAS and CMS analysis of the
LHC pp collision data at $\sqrt{s}$=7 and 8 TeV J. High Energy Phys. JHEP08 (2016) 045 arXiv:
1606.02266v2 [hep-exp].
\bibitem{13}Dolan M.J, Englert C, Spannowsky M 2012 Higgs self-coupling measurements at the LHC,
J. High Energy Phys. JHEP10(2012)112 arXiv:1206.5001v2 [hep-ph]
\bibitem{14}Zurita J 2017 Di-Higgs production at the LHC and beyond Proc. The 5 th LHCP Conf.
(Shanghai) arXiv:1708.00892 [hep-ph]
\end{thebibliography}
\end{document}